\documentclass[12pt]{iopart}

\usepackage{graphicx}
\usepackage{caption}
\usepackage{subcaption}
\usepackage{braket}
\usepackage{cite}
\usepackage{longtable}
\usepackage{amsfonts}
\usepackage[dvipsnames]{xcolor}
\usepackage{sidecap,tikz, multirow,subcaption} 
\usepackage{braket}
\expandafter\let\csname equation*\endcsname\relax
\expandafter\let\csname endequation*\endcsname\relax
\usepackage{amsmath}
\usepackage{array}
\usepackage{xcolor}
\usepackage[breaklinks=true,colorlinks=true,linkcolor=blue,urlcolor=blue,citecolor=blue]{hyperref}
\usepackage{bm,dsfont}
\usepackage[sort&compress,numbers]{natbib}

\begin{document}

\title[]{Nonlinear evolution of a cold non-relativistic electron-ion plasma with an arbitrary initial density profile: A phase mixing perspective}
\author{Subhasish Bag}
\address{Department of Physics, Indian Institute of Technology Delhi (IIT Delhi), New Delhi 110016, India}
\ead{subhasishbag95@gmail.com}

\author{Vikrant Saxena}
\address{Department of Physics, Indian Institute of Technology Delhi (IIT Delhi), New Delhi 110016, India}
\ead{vsaxena@physics.iitd.ac.in}




\begin{abstract}

Using a perturbative approach, an evolution equation for the space charge density, correct up to the third order, is deduced for arbitrary initial density profiles of the electron and ion fluids in a cold nonrelativistic plasma. The evolution equation is solved to reproduce known results pertaining to the phase mixing time in the immobile ion limit as well as in the case of mobile ions, for a homogeneous plasma as well as for a plasma with a periodic inhomogeneity. The case of non-periodic plasma inhomogeneity, as in a finite-size plasma, is also discussed and some insights are given which are well supported by fluid simulation observations.

 \end{abstract}

\section{Introduction:}
The physics of nonlinear plasma waves and oscillations has been well studied for fundamental interest as well as for understanding their role in different plasma applications such as charged particle acceleration\cite{tajima1979laser, malka2002electron, modena1995electron, faure2006controlled, rechatin2009controlling}, fast ignition\cite{tabak1994ignition}, particle heating\cite{pukhov1996relativistic}, radiation sources\cite{anand2007laser}, etc. 
It is well established that plasma oscillations, in a cold homogeneous plasma, have an amplitude  limit\cite{dawson1959nonlinear, davidson1968nonlinear} beyond which they break within a period. Davidson\cite{davidson1968nonlinear} obtained the wave-breaking limit of the initial perturbation amplitude, i.e., $max(n_e(x,0)-n_0)/n_0=0.5$. where $n_e(x,0)$ is the initial electron density profile and $n_0$ is the background ion density. The electric field becomes multi-valued at this amplitude destroying the oscillations \cite{dawson1959nonlinear}. 
Mori et al.\cite{mori1990wavebreaking} pointed out that this breaking happens when the fluid velocity and the phase velocity of oscillation become equal. The discrepancy between Dawson’s\cite{dawson1959nonlinear} and Davidson-Schram's\cite{davidson1968nonlinear} limits was resolved by Trines \cite{trines2009wave}. It was later shown by Verma et al.\cite{verma2011nonlinear} that the wave-breaking limit can be extended by adding a very small perturbation to the 2nd order mode coexisting with the fundamental mode. It was verified with a particle-in-cell simulation. Although these observations have been made for homogeneous plasma backgrounds, the inhomogeneous plasma background has also been well studied in this context.

 In an inhomogeneous plasma, the plasma frequency depends on the spatial coordinates. This leads to wave-breaking even below the wave-breaking amplitude \cite{dawson1959nonlinear}. Due to the plasma inhomogeneity, the neighboring oscillating electrons gradually go out of phase and break the coherent motion, satisfying the Dawson limit\cite{dawson1959nonlinear}. During this process, an energy transfer takes place from the initial long wavelength mode to higher order short wavelength modes, and the wave-breaking happens. This phenomenon is known as phase mixing, which is responsible for the breaking of waves (wave breaking via phase mixing) below the wave-breaking limit, and the time around which breaking occurs is called the phase-mixing time. Dawson estimated the phase mixing time due to the background inhomogeneity, which is $t\approx\frac{\pi}{2(d\omega_p/dx)X}$ where $\omega_{p}(x)=[4\pi{n}_i(x)e^2/m]^{1/2}$.

In this context, the mode coupling effect has been studied by Kaw et al. in \cite{kaw1973quasiresonant} where the time scale for energy transfer from higher to shorter wavelengths, in the sinusoidal ion background, has been estimated as $t \approx \frac{2}{\epsilon\omega_{p0}}$; $\epsilon$ being the amplitude of the background inhomogeneity. The exact solution of electron dynamics for the fixed periodic ion background has been explored by Infeld et al.\cite{infeld1989ion}. In a later work, Infeld et al.\cite{infeld1990langmuir} extended this study to a special background density profile with a local inhomogeneity (density cavity or density hill). Later, Karmakar et al.\cite{karmakar2018phase} discussed the effects of finite amplitude electron density perturbation on phase-mixing with inhomogeneous ion background. The phase mixing process is also possible for the case having the relativistic variation of electron mass \cite{infeld1989relativistic,sengupta2009phase,sengupta2011phase,sengupta2014breaking}. The characteristic electron plasma frequency acquires a spatial dependency causing neighboring oscillations to gradually move out of phase, and consequently, the wave breaks.

For the above-mentioned investigations, the ions were assumed to be non-responsive. If the ions dynamics is included, the background inhomogeneity is inherent to the plasma evolution. In the cold plasma limit, the effect of mobile ions has been    extensively investigated\cite{nappi1991modification,stewart1993nonlinear,sengupta1999phase}.
Stewart\cite{stewart1993nonlinear} extended it to the case of an electron-positron plasma. Later, Sengupta et al.\cite{sengupta1999phase} performed an analytical investigation incorporating the evolution of background ion density and calculating the approximated phase mixing time. They verified the analytically estimated phase mixing time using a 1D PIC code. Later, Verma\cite{verma2011nonlinear1} showed that no phase mixing takes place for a particular case of nonlinear electron-ion traveling waves. This happens because of the absence of ponderomotive force and zero frequency mode. They also demonstrated that cold plasma BGK waves\cite{albritton1975relation} also undergo the phase mixing process when ion dynamics is included and obtained the phase mixing time scale as $t_{mix}\approx{(1+\mu)/2\delta\mu}$. This was verified in their 1D PIC simulations. Here $\delta$ is the initial perturbation amplitude, and $\mu$ is the electron-ion mass ratio.

On a different note, the study of finite-size plasmas has attracted varied interest\cite{anand2007laser,bag2024fluid}, lately. The dynamical behaviour of a finite-size plasma differs considerably from a periodic plasma, mainly due to the presence of strong density gradients. 
Our present article addresses the phase mixing phenomenon in a cold, non-relativistic plasma for a general charge density profile to start with. It highlights how the initial plasma density profile affects the phase mixing process. The investigation is based on a two-fluid cold plasma model considering a finite electron-to-ion mass ratio. In particular, We have performed perturbation analysis to obtain the evolution equation of the charge density, which gives an estimate of the phase-mixing time in most of the cases whereas gives important insights for the other cases where estimation of phase mixing time is analytically intractable.
We also provide a comparison of the estimated phase mixing time with the energy damping time observed in our fluid simulations\cite{bag2024fluid}. The plasma evolution post phase-mixing induced wave-breaking is beyond the limitations of the fluid approach and requires a kinetic treatment, which is out of scope here.  

The article has been organized in the following manner. Section (\ref{2s}) presents the one-dimensional non-relativistic cold plasma model equations using a two-fluid approach. Section (\ref{3s}) provides an analytical treatment of the time evolution of density by perturbatively solving the two fluid cold plasma equations along with the electrostatic Poisson equation. In section (\ref{4s}), we show phase mixing damping time estimated with the analytical approaches for periodic inhomogeneous density profiles for both electron and ion fluids and then we discuss different limiting cases. The dynamics for finite-size plasma has been analyzed in section (\ref{5s}). In section (\ref{6s}) our fluid simulation results addressing the total energy evolution for different cases are presented. In particular, a comparison is drawn between the onset time of energy damping in all the relevant cases and the estimated phase mixing time discussed in sections (\ref{4s} and \ref{5s}). We finally conclude our work in section (\ref{7s}).

\section{One dimensional two-fluid cold plasma model}{\label{2s}}
We consider a two-fluid plasma model for cold, collisionless, and unmagnetized plasma. The one-dimensional two-fluid equations of our model are,
\begin{equation} \label{eq:1}
    {\frac{\partial{n_e}}{\partial{t}}}+{\frac{\partial{(n_eu_e)}}{\partial{x}}}=0 
\end{equation}
\begin{equation} \label{eq:2}
    {\frac{\partial{n_i}}{\partial{t}}}+{\frac{\partial{(n_iu_i)}}{\partial{x}}}=0 
\end{equation}
\begin{equation} \label{eq:3}
    {\frac{\partial{u_e}}{\partial{t}}}+u_e{\frac{\partial{u_e}}{\partial{x}}}=-E
\end{equation}
\begin{equation} \label{eq:4}
    {\frac{\partial{u_i}}{\partial{t}}}+u_i{\frac{\partial{u_i}}{\partial{x}}}=\mu{E}
\end{equation}
\begin{equation} \label{eq:5}
    \frac{\partial {E}}{\partial {x}}=({n_i}-{n_e})
\end{equation}

The symbols $n$ and $u$ are the fluid density and fluid velocity. The subscripts $e$ and $i$ represent the electron and ion fluids, respectively. E stands for the electrostatic field, and $\mu$ (=$\frac{m_e}{m_i}$) represents the mass ratio of electron and ion. Equations (\ref{eq:1}) and (\ref{eq:2}) are the electron and ion fluid continuity equations, respectively, while equations (\ref{eq:3}) and (\ref{eq:4}) respectively express the electron and ion fluid momentum equations. Equation (\ref{eq:5}) is the one-dimensional form of the electrostatic Gauss equation. Here, the equations are already normalized as per the following table.

\begin{table}[h]
    \centering
$$\begin{tabular}{|c|c|}
\hline
Variables  &{Normalized by}    \\ \hline
Number Density (n) &{ Maximum Density ($n_0$)}  \\ 
Time (t) &{$\omega_{p}^{-1}$ where, $\omega_{p}=\sqrt{n_0e^2/{m_e\epsilon_0}}$}  \\ 
Electric Field (E) &Maximum Initial Electric Eield at the edge (${E_{max}}$)  \\
Fluid Velocity ({$u_e,u_i$}) & Maximum Electron Velocity ($v_{max}=e E_{max}/m \omega_{pe}$)  \\
Position ({x}) &{($v_{max}/{\omega_{p}}$)}  \\  \hline
\end{tabular}$$\\
    \caption{Normalized variables}
    \label{tab1}
\end{table}

\section{Dynamical Evolution of Perturbed Plasmas}{\label{3s}}
To study the electron-ion evolution dynamics in a finite-size plasma, we follow the perturbative approach used by Sengupta et al.\cite{sengupta1999phase} for the periodic plasma case. The equations (\ref{eq:1}-\ref{eq:5}) are converted into perturbation equations. $\delta{n}_i$ and $\delta{n}_e$ are the density perturbations in ion and electron densities, respectively. We introduce the variables $\delta{n}_s$ and $\delta{n}_d$, which are the sum and difference of the electron and ion density perturbations, respectively, so that we have, $\delta{n}_d=\delta{n}_i-\delta{n}_e=n_i-n_e$ and
$\delta{n}_s=\delta{n}_i+\delta{n}_e=n_i+n_e-2$. Similarly, we define two velocity variables, $v,$ and $V$, as $v=v_i-v_e$ and $V=v_i+v_e$, respectively. The new set of equations in terms of the new variables are as follows.

\begin{equation} \label{eq:6}
    {\frac{\partial{\delta{n}_d}}{\partial{t}}}+{\frac{\partial}{\partial{x}}\left[v+\frac{\delta{n}_dV+\delta{n}_sv}{2}\right]}=0 
\end{equation}
\begin{equation} \label{eq:7}
    {\frac{\partial{\delta{n}_s}}{\partial{t}}}+{\frac{\partial}{\partial{x}}\left[V+\frac{\delta{n}_sV+\delta{n}_dv}{2}\right]}=0 
\end{equation}
\begin{equation} \label{eq:8}
    {\frac{\partial{v}}{\partial{t}}}+{\frac{\partial}{\partial{x}}\left[\frac{vV}{2}\right]}=(1+\mu)E
\end{equation}
\begin{equation} \label{eq:9}
    {\frac{\partial{V}}{\partial{t}}}+{\frac{\partial}{\partial{x}}\left[\frac{v^2+V^2}{4}\right]}=-(1-\mu)E
\end{equation}
\begin{equation} \label{eq:10}
    \frac{\partial {E}}{\partial {x}}=\delta{n}_d
\end{equation}

Now, to solve these equations (\ref{eq:6}-\ref{eq:10}), we use the perturbation approach. We consider the variable `$S$' in terms of perturbation expressed as $S=S^{(1)}+S^{(2)}+S^{(3)}+...$. In our case, $S$ is replaced by the plasma variables, $\delta{n}_s$, $\delta{n}_d$, $v$, $V$, and $E$.  After considering the 1st order correction, we arrive at the following set of equations (11-15).

\begin{equation} \label{eq:11}
    {\frac{\partial{\delta{n}_d^{(1)}}}{\partial{t}}}+\frac{\partial{v^{(1)}}}{\partial{x}}=0 
\end{equation}
\begin{equation} \label{eq:12}
    {\frac{\partial{\delta{n}_s^{(1)}}}{\partial{t}}}+\frac{\partial{V^{(1)}}}{\partial{x}}=0 
\end{equation}
\begin{equation} \label{eq:13}
    {\frac{\partial{v^{(1)}}}{\partial{t}}}=(1+\mu)E^{(1)}
\end{equation}
\begin{equation} \label{eq:14}
    {\frac{\partial{V}^{(1)}}{\partial{t}}}=-(1-\mu)E^{(1)}
\end{equation}
\begin{equation} \label{eq:15}
    \frac{\partial {E^{(1)}}}{\partial {x}}=\delta{n}_d^{(1)}
\end{equation}

Initially, we consider only a slight difference between the electron and ion density profiles such that the initial perturbation amplitude remains in the linear regime\cite{dawson1959nonlinear,davidson1968nonlinear}.

Now, if we combine equations (\ref{eq:11}-\ref{eq:15}), we obtain the following equation for $\delta{n}_d$,
\begin{equation}  \label{eq:16}
    \frac{\partial^2\delta{n}_d^{(1)}}{\partial{t}^2}+\omega_{p}^2\delta{n}_d^{(1)}=0
\end{equation}
So, with initial conditions $v_e(x,0)=v_i(x,0)=0$, the 1st order solutions are
\begin{equation}  \label{eq:17}
    \delta{n}_d^{(1)}=p(x)cos{\omega_{p}t}
\end{equation}

\begin{equation} \label{eq:18}
    v^{(1)}=\omega_{p}Sin\omega_pt\int_0^x{p(x)}dx
\end{equation}

\begin{equation} \label{eq:19}
    E^{(1)}=cos\omega_pt\int_0^x{p(x)}dx
\end{equation}

\begin{equation} \label{eq:20}
    V^{(1)}=-(1-\mu)\int_0^x{p(x)}dx\frac{Sin\omega_p{t}}{\omega_{p}}
\end{equation}

\begin{equation} \label{eq:21}
    \delta{n}_s^{(1)}=-\frac{1-\mu}{1+\mu}p(x)[cos\omega_p{t}-1]+q(x)-2
\end{equation}

where $\omega_{pe}^2=1+\mu$, $p(x)=n_i(x,0)-n_e(x,0)$, $q(x)=n_i(x,0)+n_e(x,0)$. We consider $p(x)$ to be any general initial perturbation, whereas $q(x)$ is the addition of the density profiles of electrons and ions.

Now, the 2nd-order components of the set of equations (6-10) are

\begin{equation} \label{eq:22}
    {\frac{\partial{\delta{n}_d^{(2)}}}{\partial{t}}}+{\frac{\partial}{\partial{x}}\left[v^{(2)}+\frac{\delta{n}_d^{(1)}V^{(1)}+\delta{n}_s^{(1)}v^{(1)}}{2}\right]}=0 
\end{equation}
\begin{equation} \label{eq:23}
    {\frac{\partial{\delta{n}_s^{(2)}}}{\partial{t}}}+{\frac{\partial}{\partial{x}}\left[V^{(2)}+\frac{\delta{n}_s^{(1)}V^{(1)}+\delta{n}_d^{(1)}v^{(1)}}{2}\right]}=0 
\end{equation}
\begin{equation} \label{eq:24}
    {\frac{\partial{v^{(2)}}}{\partial{t}}}+{\frac{\partial}{\partial{x}}\left[\frac{v^{(1)}V^{(1)}}{2}\right]}=(1+\mu)E^{(2)}
\end{equation}
\begin{equation} \label{eq:25}
    {\frac{\partial{V^{(2)}}}{\partial{t}}}+{\frac{\partial}{\partial{x}}\left[\frac{{v^{(1)}}^2+{V^{(1)}}^2}{4}\right]}=-(1-\mu)E^{(2)}
\end{equation}
\begin{equation} \label{eq:26}
    \frac{\partial {E^{(2)}}}{\partial {x}}=\delta{n}_d^{(2)}
\end{equation}

After solving these second order equations (\ref{eq:24}-\ref{eq:28}), the solutions can be expressed as

\begin{equation}\label{eq:27}
\begin{aligned}
    \delta{n}_s^{(2)}=\\
    & A_{c1}\left[\frac{\mu}{1-\mu^2}\frac{t^2}{2}+\frac{1-\mu}{(1+\mu)^2}\frac{\omega_ptsin\omega_pt}{4}-\frac{1-\mu}{1+\mu}\frac{cos\omega_pt}{\omega_p^2}+\frac{cos2\omega_pt}{8\omega_p^2}\left(\frac{3(1+\mu^2)}{1-\mu^2}+\frac{1-\mu}{1+\mu}\right)\right]\\
    &+B_{c1}\left[\frac{\omega_pt}{2\omega_p^2}sin\omega_pt \right]\frac{1-\mu}{1+\mu}+A_{c1}\left[\frac{1-\mu}{(1+\mu)^2}-\frac{1}{8\omega_p^2}\left(\frac{3(1+\mu^2)}{1-\mu^2}+\frac{1-\mu}{1+\mu}\right)\right]
\end{aligned}    
\end{equation}

\begin{equation}\label{eq:28}
    \delta{n}_d^{(2)}=\frac{A_{c1}}{\omega_{p}^2}cos\omega_{p}t-\frac{A_{c1}}{2\omega_{p}^2}\left(1+cos2\omega_{p}t+\frac{\omega_ptsin\omega_pt}{2}\right)-\frac{B_{c1}}{2\omega_p}{tsin{\omega_{p}t}}
\end{equation}

\begin{equation}\label{eq:29}
\begin{aligned}
    v^{(2)}=\\
    &-\frac{1}{2}\left[-p(x)cos\omega_pt(1-\mu)\left(\int{p(x)}dx\right)\frac{sin\omega_pt}{\omega_p}+\left\{\left(\frac{1-\mu}{1+\mu}\right)p(x)(1-cos\omega_pt)+q(x)-2\right\} \right.\\
    &\left. \omega_p\left(\int{p(x)dx}\right)sin\omega_pt \right]-\int{A_{c1}dx}\left[-\frac{sin{\omega_pt}}{\omega_p}+\frac{sin{2\omega_pt}}{\omega_p}-\frac{sin{\omega_pt}}{4\omega_p}-\frac{tcos{\omega_pt}}{4} \right]   \\
    & +\int B_{c1}dx\left[\frac{sin\omega_pt}{2\omega_p}+\frac{tcos\omega_pt}{2} \right]   
\end{aligned} 
\end{equation}

\begin{equation}\label{eq:30}
    {E}^{(2)}=\int A_{c1}dx\left[\frac{cos{\omega_pt}}{\omega_p^2}-\frac{1}{2\omega_p^2}-\frac{cos2\omega_pt}{2\omega_p^2}-\frac{tsin\omega_pt}{4\omega_p}\right]-\int{B_{c1}dx\frac{tsin\omega_pt}{2\omega_p}}
\end{equation}

\begin{equation}\label{eq:31}
\begin{aligned}
        {V}^{(2)}=\\
    &\frac{1+\mu^2}{1+\mu}p(x)\int{p(x)dx}\left(\frac{sin2\omega_pt}{4\omega_p}-\frac{t}{2}\right)-\frac{1-\mu}{1+\mu}\int A_{c1}dx\left[\frac{sin\omega_pt}{\omega_p}-\frac{t}{2}-\frac{sin2\omega_pt}{4\omega_p} \right.\\
    &\left. +\frac{tcos\omega_pt}{4}-\frac{sin\omega_pt}{4\omega_p} \right]-\frac{1-\mu}{1+\mu}\int B_{c1}dx\left[-\frac{tcos\omega_pt}{2}+\frac{sin\omega_pt}{2\omega_p} \right]
\end{aligned}
\end{equation}

where,
    $$A_{c1}=(1-\mu)(p(x))^2+(1-\mu)\left(\int_0^x{p(x)dx}\right)\frac{\partial{p(x)}}{\partial{x}}$$
    $$B_{c1}=\left[(q(x)-2)p(x)+\left(\int_0^x{p(x)dx}\right)\frac{\partial{q(x)}}{\partial{x}}\right]\frac{\omega_p^2}{2}$$

The second order term $\delta{n}_d^{(2)}$ explains the higher order physics, which helps in estimating the wave-breaking or phase-mixing time. These second-order solutions indicate the generation of higher harmonics in the space and time domain. One can also predict the bunching of the plasma particles in space with the harmonics. It can be inferred that phase mixing or bunching of plasma particles happens faster in the case of mobile ions than in the immobile ions case. We would like to point out that most of the previous studies have been performed for the periodic (infinite) plasma profiles, that too for a specific case. In the present work, we perform the perturbation analysis for an arbitrary initial plasma density profile and provide insights for the phase mixing induced wave breaking process. To the best of our knowledge, this is the first such analysis reported for the phase mixing for a general plasma density profile.

Now, if we write the equation for $\delta{n}_d$ correct up to third order, we get

\begin{equation}\label{eq:32}
    \frac{\partial^2}{\partial{t^2}}\delta{n}_d+\omega_p^2\left[1+\frac{1}{2}(\delta{n}_s^{(1)}+\delta{n}_s^{(2)})\right]\delta{n_d}\approx{0}
\end{equation}

Considering only the leading order secular terms from $\delta{n}_s^{(1)}$ and $\delta{n}_s^{(2)}$ (in fact, there is no leading order secular term in $\delta{n}_s^{(1)}$), the equation (\ref{eq:32}) becomes
\begin{equation}\label{eq:33}
    \frac{\partial^2}{\partial{t^2}}\delta{n}_d+\omega_p^2\left[1+\frac{\mu{A_{c1}}}{4(1-\mu^2)}{t}^2\right]\delta{n_d}\approx{0}
\end{equation}
The WKB solution of this equation (\ref{eq:32}) with the initial condition $\delta{n}_d=p(x)=n_i(x,0)-n_e(x,0)$ and $\frac{\partial{\delta{n}_d}}{\partial{t}}=0$ will be
\begin{equation}\label{eq:34}
    \delta{n_d}\approx{p(x)}cos\left[\omega_pt\left(1+\frac{t^2}{24}\frac{\mu{A_{c1}}}{(1-\mu^2)}\right)\right]
\end{equation}

The solution of $\delta{n}_d$ in equation (\ref{eq:34}), correct up to 2nd order, incorporates the effect of the phase mixing for a general one-dimensional profile of initial charge density. Based on the density profiles cold-collisionless plasmas can be broadly categorized into periodic and non-periodic or finite-size plasmas. In sections (\ref{4s}) and (\ref{5s}), the phase mixing analysis is presented for periodic and finite plasmas, respectively. Some known results are recovered for periodic plasmas before we discuss the case of finite plasmas.

\section{Nonlinear evolution of a periodic plasma}\label{4s}
We consider a two-fluid model assuming self-consistent electron and ion response. Initially, electrons and ions have sinusoidal density perturbation expressed as $n_e(x,0)=1+\delta_ecos(kx)$ and $n_i(x,0)=1+\delta_icos(kx)$. Here $\delta_i$ and $\delta_e$ are the ion and electron perturbation amplitudes, respectively, below the wave-breaking limit \cite{dawson1959nonlinear,davidson1968nonlinear}.
For such profiles, the expression for (\ref{eq:34}) becomes
\begin{equation}\label{eq:44}
    \delta{n}_d\approx(\delta_i-\delta_e){cos(kx)}cos\left[\omega_pt\left(1+\frac{t^2}{24}\frac{\mu{(\delta_e-\delta_i)^2}}{(1+\mu)}cos(2kx)\right)\right]
\end{equation}
Equation (\ref{eq:44}) can be expanded as
\begin{equation}\label{eq:45}
    \delta{n_d}\approx(\delta_i-\delta_e){cos(kx)}\sum_{n=-\infty}^{n=\infty}cos(\omega_pt+\frac{n\pi}{2}-2nkx)\times{J_n}\left(\frac{(\delta_e-\delta_i)^2t^3\mu}{24\sqrt{1+\mu}}\right)
\end{equation}
From equation (\ref{eq:45}), the phase mixing time can be estimated, using similar arguments as discussed in \cite{sengupta1999phase}, to be
\begin{equation}\label{eq:46}    
    \omega_pt_{mix}\approx\left[\frac{24\sqrt{1+\mu}}{(\delta_e-\delta_i)^2\mu}\right]^{1/3}
\end{equation}
In the remaining portion of this section, different cases are discussed, providing phase mixing time estimates, while also identifying the signatures of phase mixing in the evolution of $\delta n_d$.

\subsection{Sinusoidal perturbation in electron density with initially homogeneous ion density}

Here, we consider the ions to have a homogeneous density with a sinusoidal perturbation in electron density, at $t=0$. So, the normalized density profile will be $n_e(x,0)=1+{\delta}coskx$ and $n_i=1$. We first analyze the plasma dynamics on a slow time scale so that ions also participate in the dynamics.\\
The solution of $\delta{n}_d$ up to third order from equation (\ref{eq:34}) will be
\begin{equation}\label{eq:41}
    \delta{n}_d\approx-\delta{cos(kx)}cos\left[\omega_pt\left(1+\frac{t^2}{24}\frac{{\delta^2}\mu}{(1+\mu)}cos(2kx)\right)\right]
\end{equation}
The equation (\ref{eq:41}) can be expanded as
\begin{equation}\label{eq:42}
    \delta{n_d}\approx-\delta{cos(kx)}\sum_{r=-\infty}^{r=\infty}cos(\omega_pt+\frac{r\pi}{2}-2rkx)\times{J_r}\left(\frac{\delta^2t^3\mu}{24\sqrt{1+\mu}}\right)
\end{equation}
From equation (\ref{eq:42}), the phase mixing time can be obtained as\cite{verma2011nonlinear}
\begin{equation}\label{eq:43}    
    \omega_pt_{mix}\approx\left[\frac{24\sqrt{1+\mu}}{\delta^2\mu}\right]^{1/3}
\end{equation}
This phase mixing time in equation (\ref{eq:43}) was earlier reported by Sengupta et al.\cite{sengupta1999phase}.

\begin{figure}[!hbt]
     \begin{subfigure}[b]{0.5\textwidth}
         \includegraphics[scale=0.41]{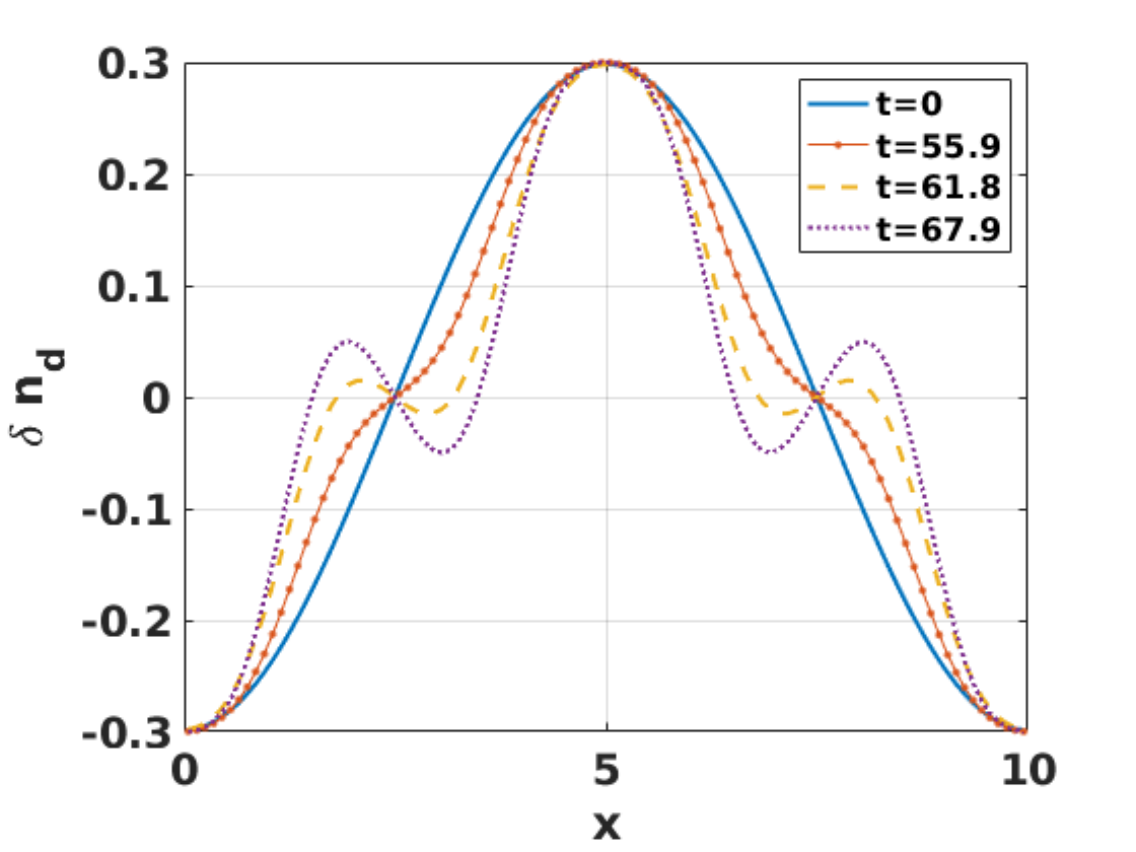}
         \caption{}
         \label{fig:a1}
     \end{subfigure}
        \begin{subfigure}[b]{0.5\textwidth}
         \includegraphics[scale=0.305]{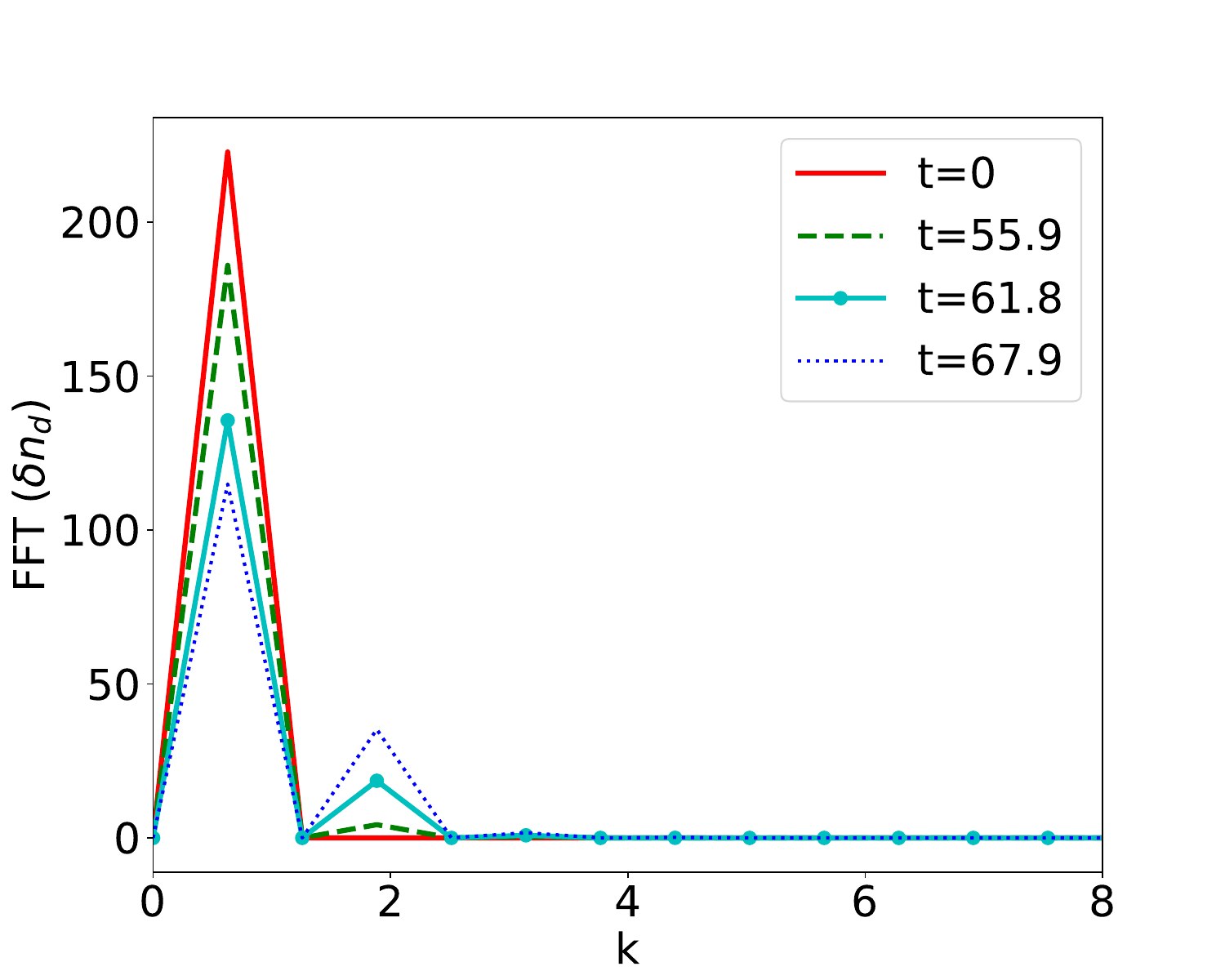}
         \caption{}
         \label{fig:b1}
     \end{subfigure}
        \caption{Evolution of the analytical solution of $\delta{n}_d$ showing the generation of higher order modes around the phase mixing damping time in (\ref{fig:a1}) for the case of periodically perturbed electron with initially homogeneous mobile ions. In fig (\ref{fig:b1}), the k-spectra shows the growth in higher order modes at the expense of initially excited modes.}
        \label{fig:ab1}
\end{figure}

The estimate of phase mixing time can also be made by looking at the time evolution of $\delta{n}_d$ (equation (\ref{eq:41})). It is shown in figure (\ref{fig:ab1}) for $\delta=0.3$. Also, the generation of higher-order spatial modes is shown in the right subplot of figure (\ref{fig:ab1}). From the figure, we note that the time of wave structure modification and that of the generation of higher modes approximately matches the analytically estimated phase-mixing time (equation (\ref{eq:43})) $\approx{64.38}$ $\omega_p^{-1}$. We therefore propose the appearance of non-reversible changes in the profile of $\delta n_d$ as an indicator for phase-mixing induced wave breaking.

 Now, if we study the plasma dynamics on a fast time scale such that ions can be assumed to be infinitely heavy ($\mu=\frac{m_e}{m_i}=0$) such that $n_i(x,t)=n_0$. The initial conditions for electrons are $n_e(x,0)=n_0(1+\delta{cos(kx))}$ and $v_e(x,0)=0$. The solution for $\delta{n}_d$ can then be written, using equations (\ref{eq:17}) and (\ref{eq:28}), as 
\begin{equation}\label{eq:36}
    \begin{aligned}
        \delta{n}_d &=\frac{n_e(x,t)-n_0}{n_0}\\
    &=\delta{cos(kx)}cos(\omega_pt)-\delta^2cos(2kx)cos(\omega_pt)\{1-cos(\omega_pt)\}+...\\
    &=\delta{n}_d^{(1)}+\delta{n}_d^{(2)}+...
    \end{aligned}
\end{equation}
Again the total solution of $\delta{n}_s$ using the solution (\ref{eq:21}) and (\ref{eq:27}) we have,
\begin{equation}\label{eq:37}
    \begin{aligned}
        \delta{n}_s &=\frac{n_i(x,t)-n_0+n_e(x,t)-n_0}{n_0}\\
        &=\frac{n_e(x,t)-n_0}{n_0}\\
    &=\delta{cos(kx)}cos(\omega_pt)-\delta^2cos(2kx)cos(\omega_pt)\{1-cos(\omega_pt)\}+...\\
    &=\delta{n}_s^{(1)}+\delta{n}_s^{(2)}+...
    \end{aligned}
\end{equation}
Considering any one expression from equations (\ref{eq:36}) or (\ref{eq:37}), we can express the evolution of electron density as
\begin{equation}\label{eq:35}
    n_e(x,t)=n_0+n_0\delta{cos(kx)}cos(\omega_pt)-n_0\delta^2cos(2kx)cos(\omega_pt)\{1-cos(\omega_pt)\}+...
\end{equation}

Similarly, using equations (\ref{eq:18}) and (\ref{eq:29}) or (\ref{eq:20}) and (\ref{eq:31}), we get the expression for $v_e(x,t)$. Again we have the electric field expression using the solutions (\ref{eq:19}) and (\ref{eq:30}).
These expressions can be reorganized in a compact form, to recover the exact solutions obtained in the seminal work of Davidson\cite{davidson1968nonlinear} as

\begin{equation}\label{eq:38}
    \begin{split}
        & v_e(x,t)=\frac{\omega_p}{k}{\delta}\sum_{r=1}^{\infty}(-1)^r\frac{2}{r\Omega(t)}J_r(r\Omega(t))sin(rkx)sin(\omega_pt)
         \\
        & E(x,t)=-\frac{m}{e}\frac{\omega_p^2}{k}{\delta}\sum_{r=1}^{\infty}(-1)^r\frac{2}{r\Omega(t)}J_r(r\Omega(t))sin(rkx)cos(\omega_pt)
         \\
         & n_e(x,t)=n_0+\frac{2n_0\delta}{\Omega(t)}\sum_{r=1}^{\infty}(-1)^rJ_r(r\Omega(t))cos(rkx)cos(\omega_pt)
    \end{split}
\end{equation}

Here $J_r$ are the Bessel function of first kind and $\Omega(t)=2\delta{sin}^2\frac{\omega_pt}{2}$. The initial perturbation amplitude $\delta$ is below the wave breaking limit as $|\delta|<\frac{1}{2}$. These results also explain the wave distortions for the nonlinear effect through the generation of higher harmonics. As there is no spatial dependency on plasma frequency in this case, no phase mixing occurs. The oscillation breaks within a period only if the initial perturbation amplitude is higher than the wave-breaking limit\cite{dawson1959nonlinear,davidson1968nonlinear}, i.e., $|\delta|<\frac{1}{2}$. Otherwise, the oscillation is sustained forever.

\subsection{Sinusoidal perturbation in the densities of electron and (mobile) ion fluids}

\begin{figure}[!hbt]
     \begin{subfigure}[b]{0.5\textwidth}
         \includegraphics[scale=0.41]{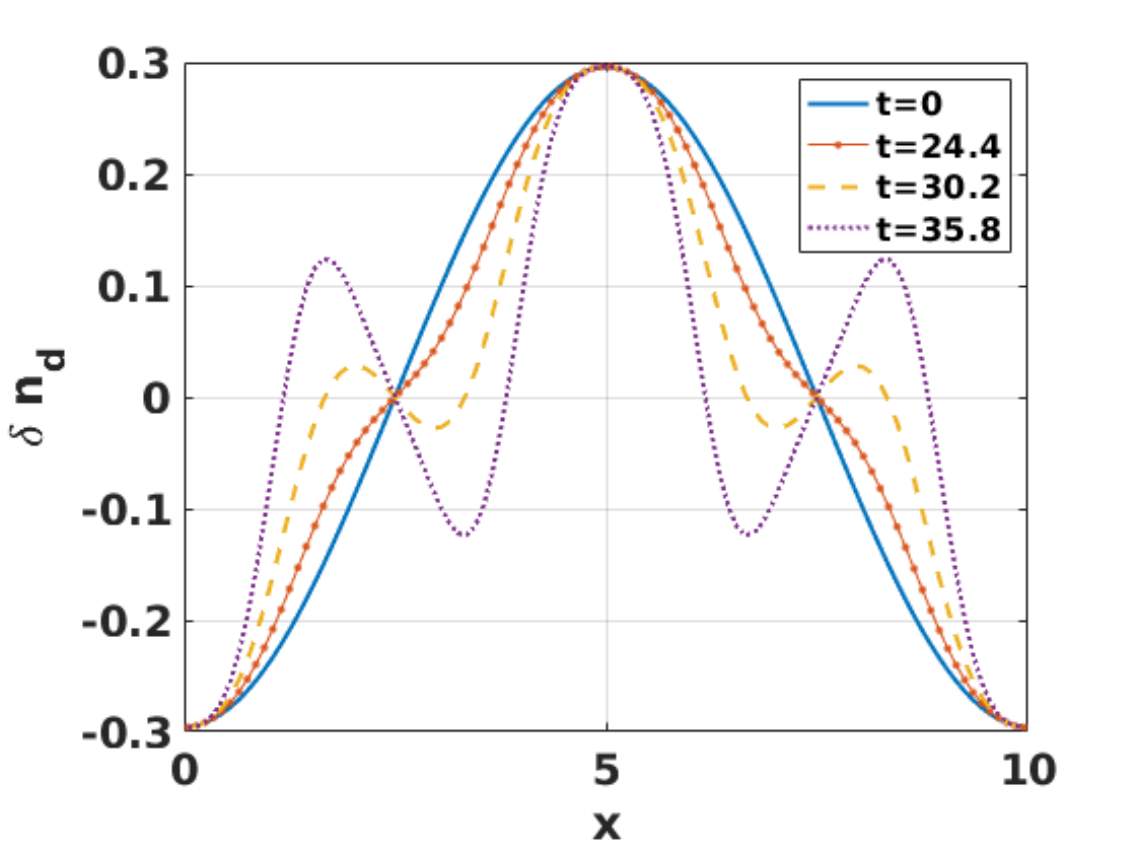}
         \caption{}
         \label{fig:a2}
     \end{subfigure}
        \begin{subfigure}[b]{0.5\textwidth}
         \includegraphics[scale=0.305]{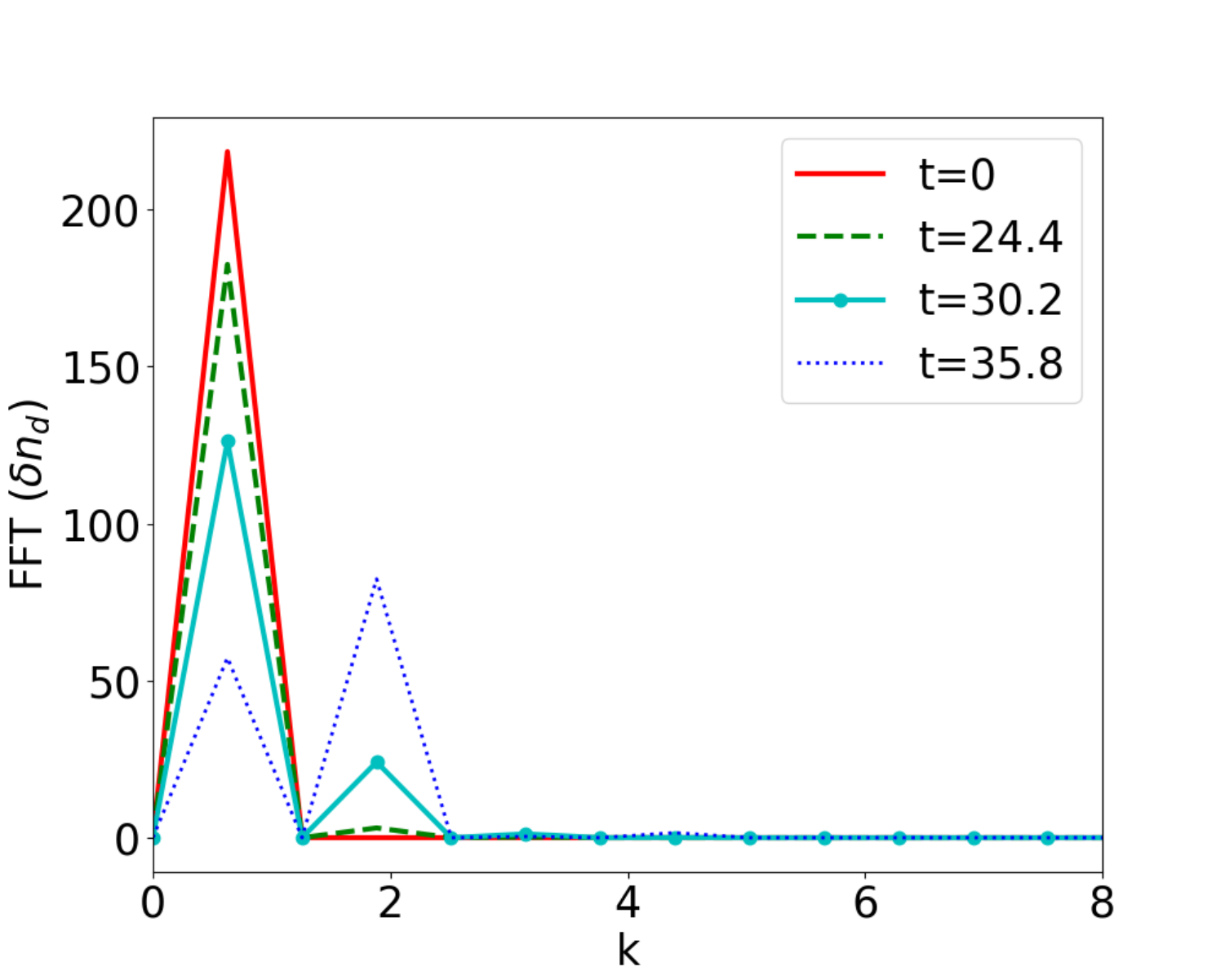}
         \caption{}
         \label{fig:b2}
     \end{subfigure}
        \caption{Evolution of the analytical solution of $\delta{n}_d$ showing the generation of higher order modes around the phase mixing time in (\ref{fig:a2}) for the case of periodically perturbed electrons and ions. In fig (\ref{fig:b2}), the k-spectra shows the energy transfer to higher-order modes.}
        \label{fig:ab2}
\end{figure}

In this case, both electrons and ions fluids are considered to have sinusoidal density perturbation\cite{verma2011nonlinear1} in such a way that $\delta_i=\mu\delta_e$ is relatively low due to the heavy ion mass. The densities are then expressed as $n_e(x,0)=1+\delta_ecos(kx)$ and $n_i(x,0)=1+\mu{\delta_e}cos(kx)$. Now, considering $\delta_e=0.3$ and $\delta_i=0.003$ (choosing $\mu=0.01$), we have shown the evolution of $\delta{n}_d$ in figures (\ref{fig:ab2}). The calculated phase mixing time from equation (\ref{eq:44}) is $\approx{30.13}$ $\omega_p^{-1}$. The evolution of $\delta{n}_d$, along with the growth of higher modes around the estimated phase mixing time, has been shown in figure (\ref{fig:ab2}). The evolution of spatial Fourier modes is shown in figure (\ref{fig:b2}) at time instants corresponding to figure (\ref{fig:a2}). This comparison strengthens our proposed diagnostic for estimating the phase mixing time through the numerical solution of $\delta{n}_d$.

\subsection{Electron plasma oscillations in an inhomogeneous static ion background}

 In this case, ions are assumed to be immobile, i.e., $\mu=0$, while having an inhomogeneous density profile as $n_i(x,t)=1+\delta{cos(kx)}$ with $\delta<1/2$. The electron fluid is considered to have a homogeneous density profile to start with, i.e., $n_e=1$ (normalized). This is a similar case as investigated by Infeld et al.\cite{infeld1989ion} and the charge density amplitude is taken below the wave breaking limit. From equations ($\ref{eq:17}$) and ($\ref{eq:21}$) we have
\begin{equation}
    \begin{aligned}
        &n_e^{(1)}(x,t)=1+\delta{cos(kx)}(1-cos\omega_pt)\\
        &n_i^{(1)}(x)=1+\delta{cos(kx)}
    \end{aligned}
\end{equation}
Again the equations ($\ref{eq:27}$) and ($\ref{eq:28}$) we have
\begin{equation}\label{eq:40}
    \begin{aligned}
        &n_e^{(2)}(x,t)=1+\delta^2{cos(2kx)}\left[\frac{tsin(\omega_pt)}{2}+cos\omega_pt\left(cos\omega_pt-1\right)\right]\\
        &n_i^{(2)}(x)=1
    \end{aligned}
\end{equation}

The expression of $n_e^{(2)}$ in equation (\ref{eq:40}) is different from equation (\ref{eq:35}), as there is an additional amplitude increasing term (secular term) with time which arises due to the initial inhomogeneity in the ion density. This secular term makes the ponderomotive force non-zero, which helps in the coupling of plasma modes\cite{kaw1973quasiresonant}. The mode coupling effect\cite{kaw1973quasiresonant} has been introduced in the solution with the initial ion inhomogeneity, which was not present in the case of homogeneous ion background\cite{davidson1968nonlinear}. This mode coupling ultimately implies the phase mixing taking place in finite time\cite{kaw1973quasiresonant,infeld1989ion,infeld1990langmuir} unlike in the case of a homogeneous immobile ion background, wherein the phase mixing time is infinite\cite{sengupta1999phase}. The calculation of phase-mixing induced wave breaking time, in this case, would need a higher order perturbation analysis which is out of scope here and may be attempted in future.

\section{Nonlinear evolution of a finite-size plasma}\label{5s}
Finally, we extend our analysis to investigate the dynamics of finite-size plasmas. For a finite-size plasma\cite{bag2024fluid,anand2007laser}, we consider the double tangent hyperbolic plasma density profile having initial perturbations which can be described by the following mathematical expressions (\ref{eq:48}) and (\ref{eq:17}) for electron and ion densities\cite{bag2024fluid}, respectively. To maintain the quasi-neutrality, the areas under the curve for electron and ion density profiles are kept the same. This has been done by adjusting the profile widths and amplitudes. The factor $\beta$ is used to maintain the quasi-neutrality.

\begin{equation}\label{eq:48}
    n_i(x,0)=tanh\left(\frac{x-a_1}{\alpha}\right)-tanh\left(\frac{x-b_1}{\alpha}\right) 
\end{equation}
\begin{equation}\label{eq:49}
    n_e(x,0)=(1 - \tilde{n})\times\left[tanh\left(\frac{x-a_2}{\alpha}\right)-tanh\left(\frac{x-b_2}{\alpha}\right)\right]
\end{equation}

In equation (\ref{eq:48}), the width is decided by the constants $a_1$ and $b_1$; $\alpha$ decides the density scale length, while the amplitude has been normalized to 1. In the electron density profile expression (\ref{eq:49}), $a_2$ and $b_2$ decide the width of the profiles, while an appropriate value of the factor $(1 - \tilde{n})$  ensures the quasi-neutrality. This gives us symmetric plasma profiles at the center of the simulation box. 


The generalized equation (\ref{eq:34}) can be written as 
\begin{equation}\label{eq:50}
    \delta{n_d}\approx{p(x)}cos\left[\omega_pt\left(1+\frac{t^2}{24}\frac{\mu{A_{c1}}}{(1-\mu^2)}\right)\right]
\end{equation}
Using the initial profiles of $n_i(x,0)$ and $n_e(x,0)$ we have
    $$\frac{\partial}{\partial{x}}p(x)=\frac{1}{\alpha}\left[\frac{1}{cosh^2(\frac{x-a_1}{\alpha})}-\frac{1}{cosh^2(\frac{x-b_1}{\alpha})}-\frac{\beta}{cosh^2(\frac{x-a_2}{\alpha})}+\frac{\beta}{cosh^2(\frac{x-b_2}{\alpha})}\right]$$
    $$\int{p(x)dx}=\alpha{log}\left[\frac{cosh\left(\frac{x-a_1}{\alpha}\right)}{cosh\left(\frac{x-b_1}{\alpha}\right)}\right]+\alpha\beta{log}\left[\frac{cosh\left(\frac{x-b_2}{\alpha}\right)}{cosh\left(\frac{x-a_2}{\alpha}\right)}\right]$$
    Where, $$A_{c1}=(1-\mu)(p(x))^2+(1-\mu)\left(\int_0^x{p(x)dx}\right)\frac{\partial{p(x)}}{\partial{x}}$$

\begin{figure}[!hbt]
     \begin{subfigure}[b]{0.5\textwidth}
         \includegraphics[scale=0.42]{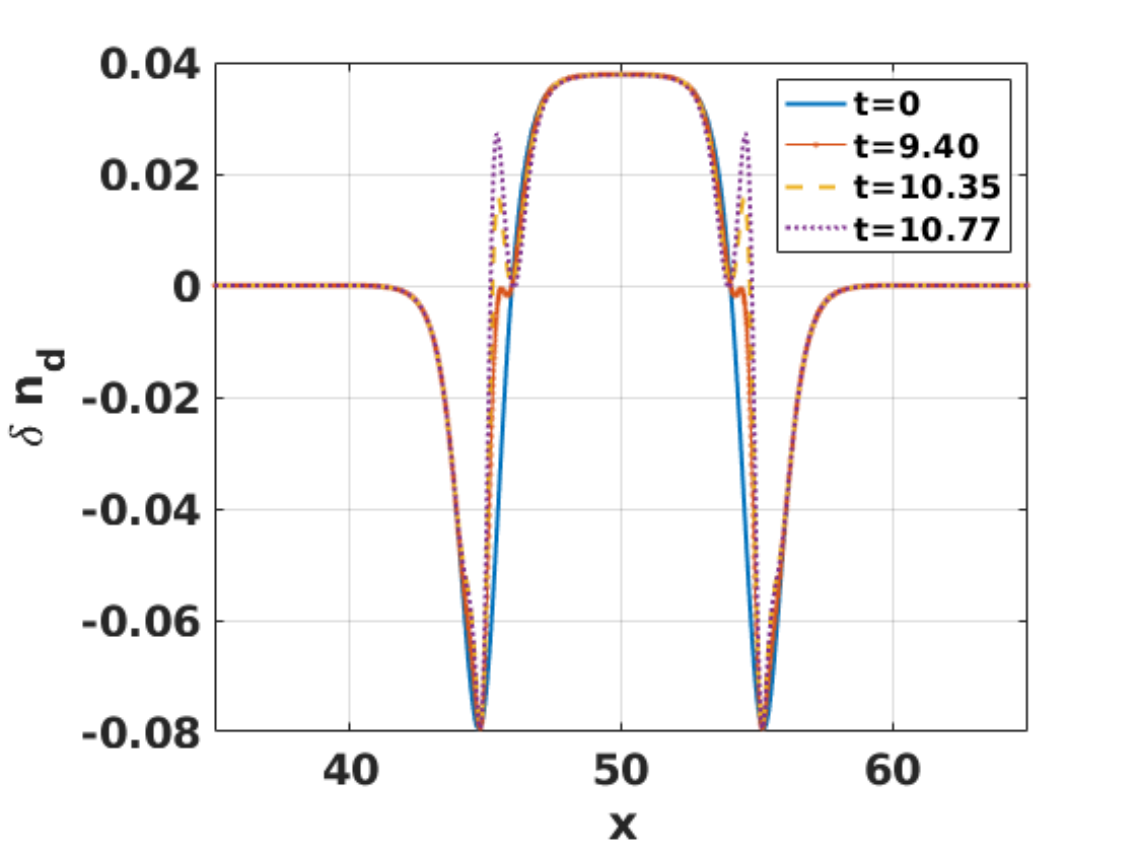}
         \caption{}
         \label{fig:a4}
     \end{subfigure}
        \begin{subfigure}[b]{0.5\textwidth}
         \includegraphics[scale=0.265]{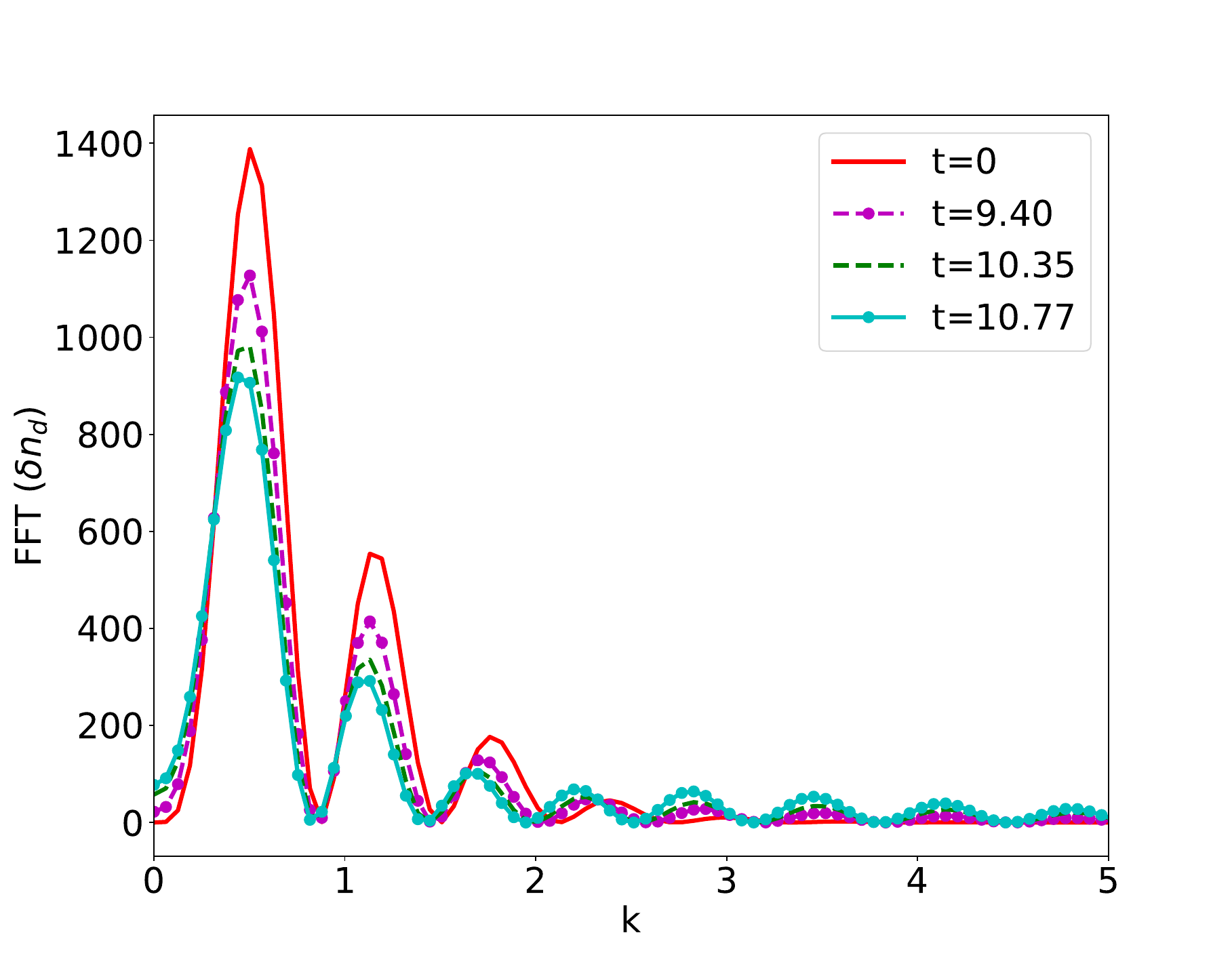}
         \caption{}
         \label{fig:b4}
     \end{subfigure}
        \caption{Evolution of the analytical solution of $\delta{n}_d$ showing the generation of higher order modes around the phase mixing damping time in (\ref{fig:a4}) for the case of a finite-size plasma. In fig (\ref{fig:b4}), the k-spectra shows the growth of the higher order modes.}
        \label{fig:ab4}
\end{figure}

    The non-zero $A_{c1}$ is responsible for the influence of the phase mixing process. As estimating the phase mixing time is not analytically tractable in this case, the phase mixing time is numerically estimated using the proposed diagnostics based on the evolution of $\delta{n}_d$ profile.

    We now consider a finite-size plasma where density profiles of electron and ion fluids are defined by equations (\ref{eq:48}) and (\ref{eq:49}), respectively. The parameter $\alpha$=1 defines the sharpness of the edge region. We have considered $a_i=44.95$, $b_i=55.05$, $a_e=44.75$, and $b_e=55.25$ for this density profile and $\Tilde{n}$ is 0.96 to ensure the integrated densities of both species are the same to satisfy the quasi-neutrality criteria. The maximum density is unity for ions, but the width and amplitude of the electron density profile are chosen to be slightly different to incorporate the initial charge imbalance. The evolution of $\delta{n_d}$ and the higher order spatial mode has been shown in figure- (\ref{fig:ab4}). From these plots, we can estimate the phase-mixing time to be approximately 10.35 $\omega_p^{-1}$.
    
To confirm this phase mixing time for the finite plasma estimated from the time evolution of $\delta{n}_d$, in the following section we look at the evolution of total energy in the corresponding two-fluid simulations and compare the phase mixing time with the onset time for energy damping in the fluid simulations.

\section{Fluid Simulation of periodic and finite size plasma}{\label{6s}}
We have performed fluid simulation for all the cases explained in sec-(\ref{4s}) and sec-(\ref{5s}) to support our proposed diagnostic for approximately estimating the phase mixing induced wave breaking time. A flux corrected scheme (FCT)\cite{boris1993lcpfct, boris1976solution} has been used to solve the set of fluid equations while a 1D Poisson solver (cyclic tridiagonal method\cite{press1992numerical}) has been employed to solve the Poisson equation connecting the electron and ion fluids.

 The total energy for the dynamical system is the sum of kinetic and potential energy and can be expressed as 
\begin{equation} \label{eq:51}
    E_{t}=E_{ke}+E_{ki}+E_{pot}
\end{equation}
Here $E_t$, $E_{ke}$, $E_{ki}$, and $E_{pot}$ are the normalized total, electron kinetic, ion kinetic, and potential energy, respectively.  
In ref (\cite{bag2024fluid}), it was shown that phase mixing-induced wave-breaking happens when the total energy starts to damp, which is also associated with the violation of the fluid model. In the following the fluid simulation results are presented addressing the time evolution of the total energy along with electron kinetic energy and the potential energy for all the cases discussed in the previous sections. 

\begin{figure}
    \centering
    \includegraphics[width=0.5\linewidth]{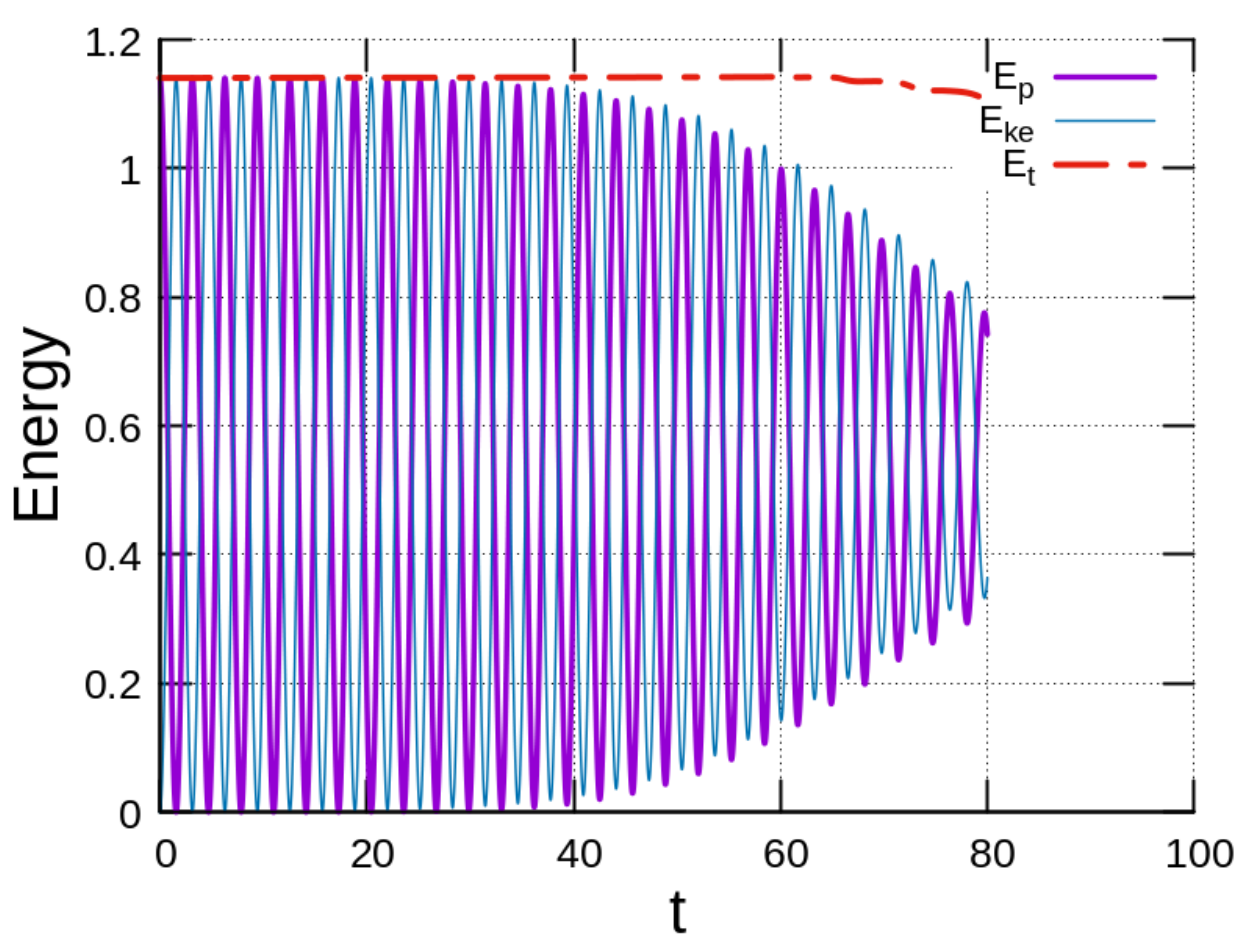}
    \caption{Total Energy for initially periodic electron density ($n_e(x,0)=1+0.3coskx$) having mobile ($\Delta=\frac{m_e}{m_i}=0.001$) homogeneous ion background where $E_p$, $E_{ke}$, $E_{t}$ are the Potential Energy, Electron Kinetic Energy, and Total Energy, respectively. Damping starts at  $\approx{64}$ $\omega_p^{-1}$.}
    \label{fig:1}
\end{figure}

As in the previous section, we first consider the case involving sinusoidal electron density $n_e(x,0)=1+0.3coskx$ and mobile ions ($\mu=\frac{m_e}{m_i}=0.001$) with a homogeneous density profile initially. The phase-mixing induced wave-breaking time($\omega_pt_{mix}$) from equation (\ref{eq:37}) is $\approx{64.38}$ $\omega_p^{-1}$. From the fluid simulations, the total energy damping is shown in figure (\ref{fig:1}), and the damping onset time is noted to be around $\approx{64}$ $\omega_p^{-1}$ which is almost same as the analytically estimated phase mixing time.

\begin{figure}[!hbt]
     \begin{subfigure}[b]{0.5\textwidth}
         \includegraphics[scale=0.375]{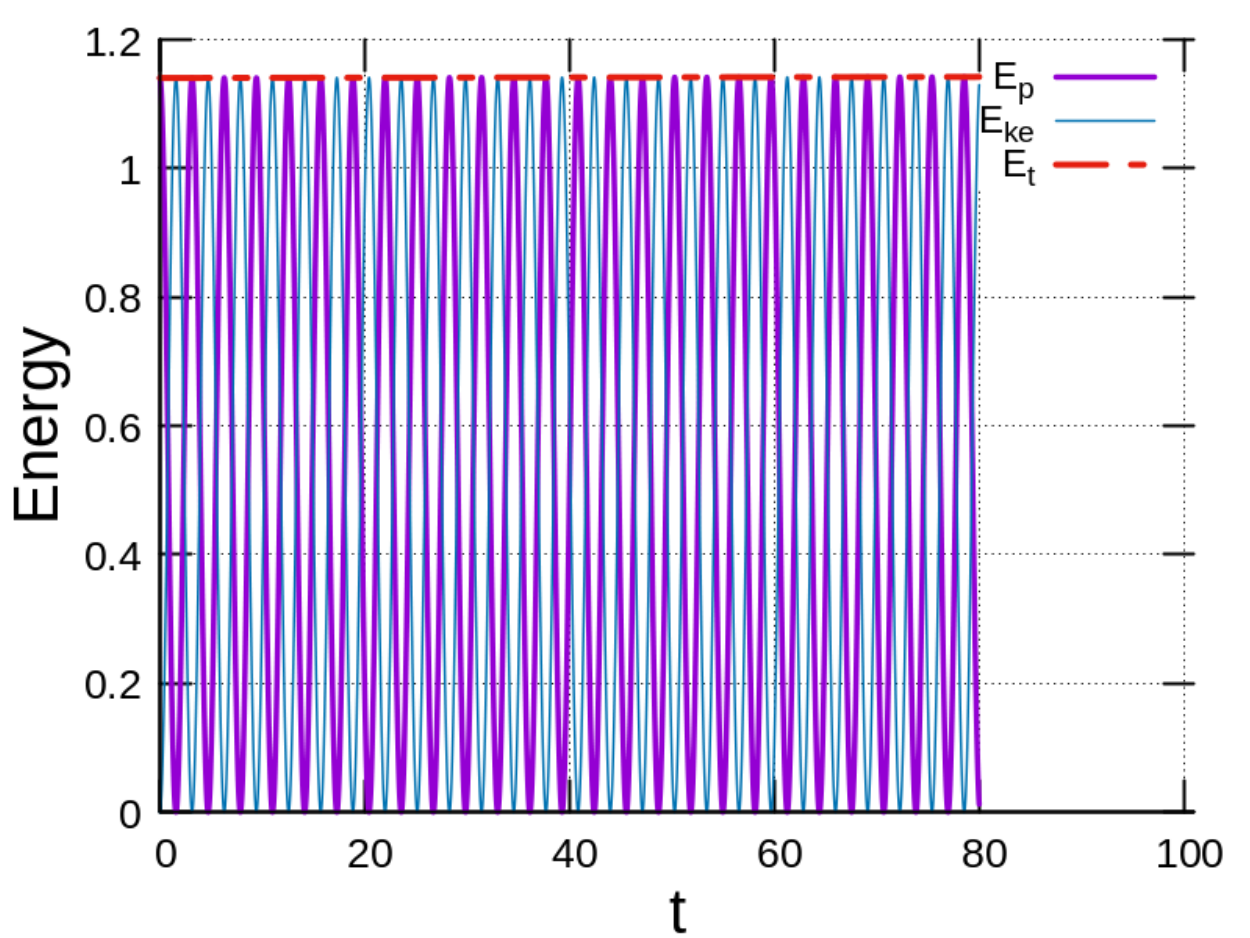}
         \caption{}
         \label{fig:2a}
     \end{subfigure}
        \begin{subfigure}[b]{0.5\textwidth}
         \includegraphics[scale=0.375]{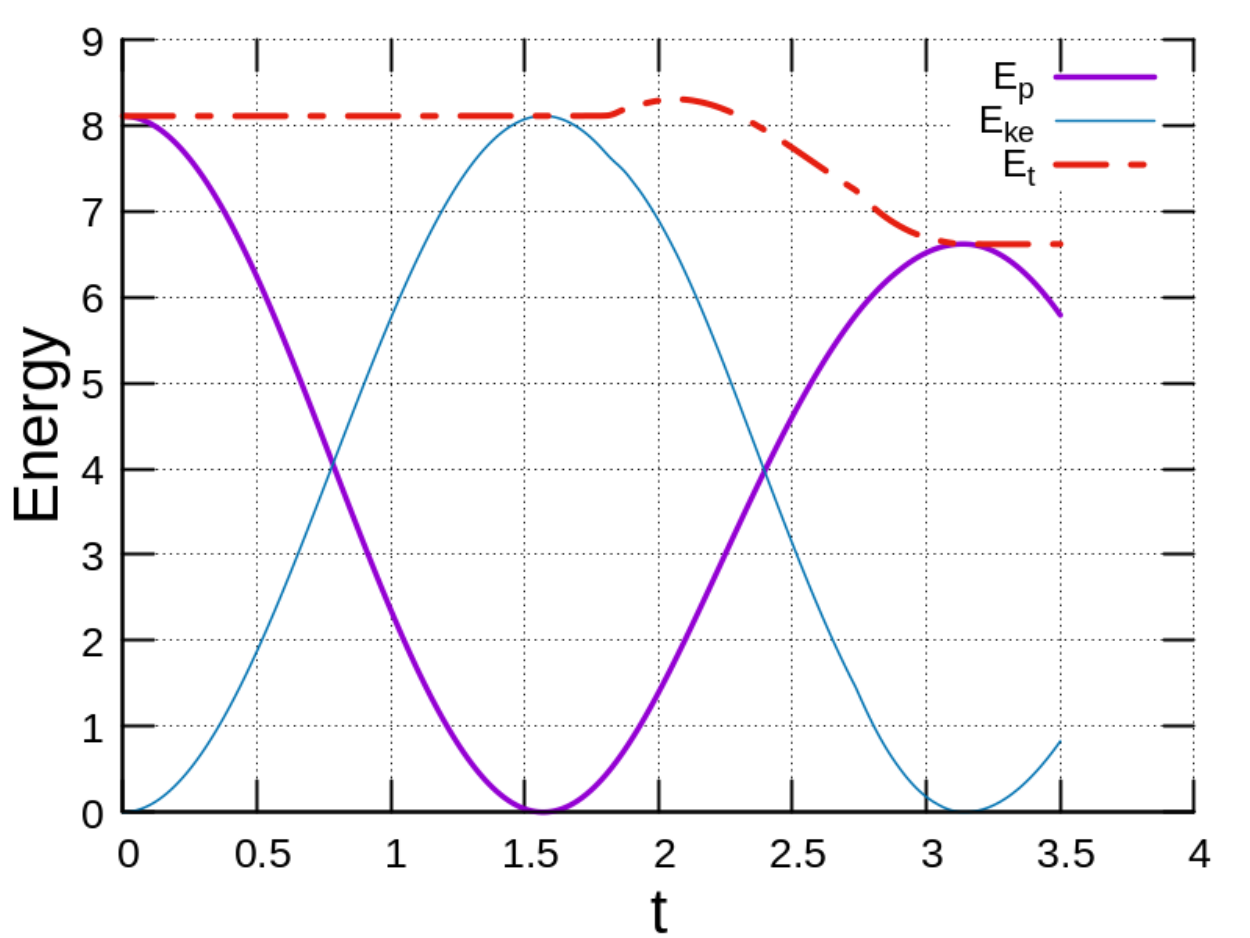}
         \caption{}
         \label{fig:2b}
     \end{subfigure}
        \caption{Total Energy for initially periodic electron density ($n_e(x,0)=1+{\delta}coskx$) having immobile ($\Delta=\frac{m_e}{m_i}=0$) homogeneous ion background where $E_p$, $E_{ke}$, $E_{t}$ are the Potential Energy, Electron Kinetic Energy, and Total Energy, respectively. Here, $\delta$ is 0.3 for figure (\ref{fig:2a}) and 0.8 for figure (\ref{fig:2b}). No wave-breaking happens for $\delta=0.3$ (as $\delta<0.5$), and breaking has been observed within a period of oscillation for $\delta=0.8$ (as $\delta>0.5$).}
        \label{fig:2}
\end{figure}

Similarly, for the case of immobile ions with homogeneous density profile we have verified in the fluid simulations that if the initial electron density perturbation amplitude ($\delta$) is 0.3 in figure (\ref{fig:2a}), there will be no breaking in the wave. When we consider the initial perturbation amplitude ($\delta$) to be 0.8, i.e., greater than the wave-breaking amplitude limit of 0.5, the breaking happens within a period (figure-\ref{fig:2b}).

\begin{figure}
    \centering
    \includegraphics[scale=0.375]{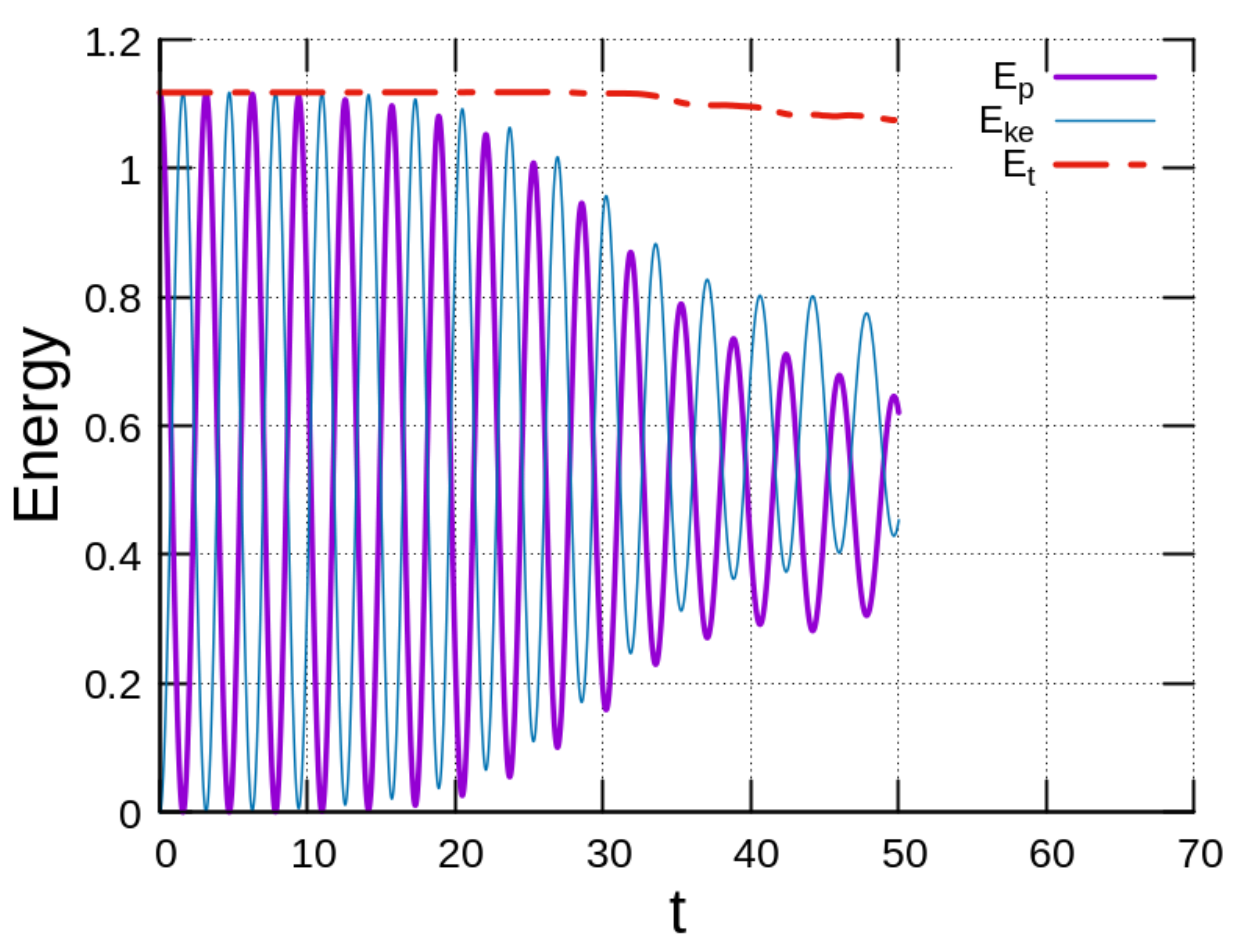}
    \caption{Total Energy for initially periodic electron density ($n_e(x,0)=1+{\delta_e}coskx$) having mobile ($\Delta=\frac{m_e}{m_i}=0.01$) inhomogeneous ion ($n_e(x,0)=1+{\delta_i}coskx$) where $E_p$, $E_{ke}$, $E_{t}$ are the Potential Energy, Electron Kinetic Energy, and Total Energy, respectively. Here, $\delta_e$ and $\delta_i$ are 0.3 and 0.003 respectively. The predicted wave-breaking time from the energy damping is $\approx{30}$ $\omega_p^{-1}$. }
    \label{fig:3a}
\end{figure}

In the case of mobile ions with an inhomogeneous density profile, given by  $n_i(x,0)=1+0.003coskx$, the phase-mixing induced wave-breaking damping time has been calculated from equation (\ref{eq:44}) as $\approx{30}$ $\omega_p^{-1}$. The total energy evolution for this case has been shown in figure (\ref{fig:3a}). Please note that the total energy damping is set in at around $\approx{28}$ $\omega_p^{-1}$ which strongly indicates the occurrence of phase mixing induced wave-breaking. 

\begin{figure}
    \centering
    \includegraphics[width=0.5\linewidth]{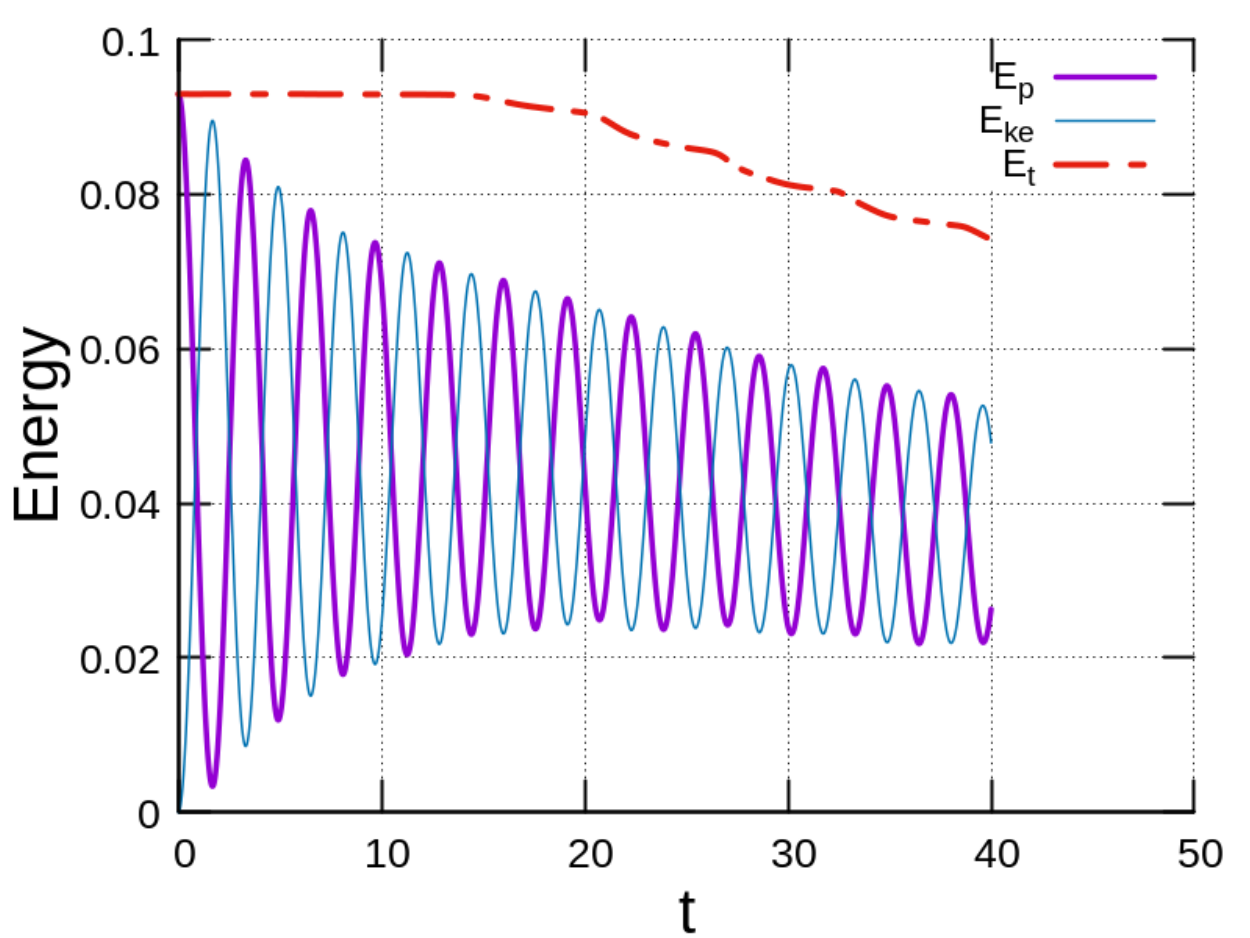}
    \caption{Total Energy for the case of a finite size plasma having mobile ($\Delta=\frac{m_e}{m_i}=0.001$) ions where $E_p$, $E_{ke}$, $E_{t}$ are the Potential Energy, Electron Kinetic Energy, and Total Energy, respectively. Damping sets in at around $\approx{13}$ $\omega_p^{-1}$.}
    \label{fig:4}
\end{figure}

All the above simulation results further support our proposed diagnostic for phase mixing induced wave-breaking in terms of irreversible node formation in the spatial profile of $\delta n_d$. Finally, the fluid simulations are performed with the initial conditions representing a finite-size plasma, as discussed in section (\ref{5s}). The resultant time evolution of the total energy is shown in figure (\ref{fig:4}). The total energy damping for the finite-size plasma considered here is observed to kick in at around 13 $\omega_p^{-1}$, which is around the phase-mixing time $10.35$ $\omega_p^{-1}$ estimated based on the evolution of $\delta n_d$. For analytically estimating the phase mixing time in the finite plasma case a higher order perturbation analysis may be helpful although is out of scope here.

\section{Conclusions}\label{7s}
We have investigated, using perturbation analysis, the phase mixing of plasma oscillations for a general initial density profile. A general evolution equation, correct up to third order, is obtained for the space charge density. Few known results for homogeneous as well as inhomogeneous initial ion density profiles are recovered for the case of periodic plasma. 
It is found that for a homogeneous static ion background, no secular term appears even in the higher-order solutions. However, in the case of an inhomogeneous static ion background, the secular term exists in the higher-order solution and is responsible for the phase mixing process. 
It is shown that the phase mixing process happens faster for the initial inhomogeneous mobile ion background than the homogeneous mobile ion case. This is because the initial electrostatic field is stronger due to the higher charge separation for the inhomogeneous case. We have proposed a new diagnostic for phase mixing based on the evolution of the $\delta n_d$ profile. This predicts the approximate phase mixing time very close to the analytically estimated phase mixing time and can be used in the cases where an analytical estimate of phase mixing time is not tractable.

The analysis is extended for a finite plasma where the initial ion as well as electron density profiles are inhomogeneous and non-periodic. Considering a double tangent hyperbolic density profile, for electron as well as ion density, we have derived the evolution equation for the higher-order component of the charge density perturbation. However, due to the non-sinusoidal nature of the charge density, the estimation of phase mixing time is not analytically tractable, unlike in the case of a periodic charge density perturbation.  
We therefore estimate the phase mixing time based on the numerical solution of the evolution equation for $\delta n_d$. Finally, the phase mixing time estimated analytically or from the numerical solution of the charge density evolution equation agrees well with the total energy-damping onset time in the corresponding two-fluid simulations.

\bibliographystyle{unsrt}
\bibliography{library}

\end{document}